\documentclass[superscriptaddress,preprint,amsmath,amssymb,aps, physrev]{revtex4-2}

\usepackage{graphicx}
\usepackage{dcolumn}
\usepackage{bm}

\usepackage{siunitx} 
\usepackage{xcolor}

\usepackage[
    colorlinks=true,
    linkcolor=blue,
    citecolor=blue,
    urlcolor=blue]{hyperref}
\usepackage[all]{hypcap}

\begin{document}

\title{First-Principles Electron-Magnon Coupling with Machine-Learning Hamiltonians: From Band Renormalization to Transport}

\author{Shixu Liu}
\thanks{These authors contributed equally.}
  \affiliation{Department of Physics, Key Laboratory of Computational Physical Sciences (Ministry of Education), Institute of Computational Physical Sciences, Fudan University, Shanghai, 200438, China}
\author{Xingding Li}
\thanks{These authors contributed equally.}
  \affiliation{Department of Physics, Key Laboratory of Computational Physical Sciences (Ministry of Education), Institute of Computational Physical Sciences, Fudan University, Shanghai, 200438, China}
\author{Haozhe Li}
\thanks{These authors contributed equally.}
  \affiliation{Department of Physics, Key Laboratory of Computational Physical Sciences (Ministry of Education), Institute of Computational Physical Sciences, Fudan University, Shanghai, 200438, China}
\author{Yang Zhong}
  \affiliation{Department of Physics, Key Laboratory of Computational Physical Sciences (Ministry of Education), Institute of Computational Physical Sciences, Fudan University, Shanghai, 200438, China}
\author{Hongjun Xiang}
\email{hxiang@fudan.edu.cn}
  \affiliation{Department of Physics, Key Laboratory of Computational Physical Sciences (Ministry of Education), Institute of Computational Physical Sciences, Fudan University, Shanghai, 200438, China}
\author{Xin-Gao Gong}
\email{xggong@fudan.edu.cn}
  \affiliation{Department of Physics, Key Laboratory of Computational Physical Sciences (Ministry of Education), Institute of Computational Physical Sciences, Fudan University, Shanghai, 200438, China}
  \affiliation{Hefei National Laboratory, Hefei, 230088, China}
\author{Ji-Hui Yang}
\email{jhyang04@fudan.edu.cn}
  \affiliation{Department of Physics, Key Laboratory of Computational Physical Sciences (Ministry of Education), Institute of Computational Physical Sciences, Fudan University, Shanghai, 200438, China}
  \affiliation{Hefei National Laboratory, Hefei, 230088, China}

\date{\today}

\begin{abstract}
    In analogy to electron-phonon coupling (EPC), electron-magnon coupling (EMC) is expected to shape electronic structure, transport, and possibly unconventional superconductivity in magnetic materials. However, unlike EPC, which is now routinely treated within first-principles frameworks, a quantitative description of EMC, especially for transport, remains elusive because of the lack of theoretical formalism. Consequently, even for elemental iron, EPC-only calculations miss both the magnitude and the $T^2$ component of resistivity. This discrepancy has long been attributed to EMC, although direct computational evidence has been lacking and the underlying transport mechanism remains unresolved. Here we develop a unified first-principles formalism for EMC in collinear magnetic systems within many-body perturbation theory, complemented by machine-learning spinful Hamiltonians that supply quantities not directly accessible from conventional first-principles methods. Our framework enables \textit{ab initio} transport calculations including EMC effects for the first time. Applied to ferromagnetic $\alpha$-Fe, our approach yields electron spectral functions consistent with previous studies. More importantly, we recover the full $T^2$ component of resistivity with a coefficient in quantitative agreement with measurement and reveal that the $T^2$ component cannot be attributed solely to EMC, as has long been assumed, but is dominated by the strong EPC-EMC interplay. Extending to antiferromagnetic K-doped $\mathrm{BaMn_2As_2}$, our method captures the ARPES-observed magnon-induced kink and a large EMC strength of $\sim 3$ comparable to experimental measurements, demonstrating the generality of the framework. Our work closes a longstanding gap in the quantitative understanding of transport in magnetic systems and provides a predictive foundation for examining magnon-mediated phenomena.
\end{abstract}

\keywords{electron-magnon interaction, transport in magnetic systems, first-principles, machine-learning}

\maketitle

Electron-boson coupling is a central theme in condensed-matter physics. Among its various manifestations, electron-phonon coupling (EPC) is one of the success stories of first-principles physics, now routinely and quantitatively captured within frameworks such as density functional perturbation theory (DFPT)~\cite{giustinoElectronphononInteractionsFirst2017}. In magnetic materials, however, electrons can couple not only to phonons but also to magnons, which are the collective spin excitations of a magnetically ordered system. Like EPC, electron-magnon coupling (EMC) underpins a wide range of phenomena, including electronic structure~\cite{yuStrongBandRenormalization2022}, charge transport~\cite{carpeneDynamicsElectronmagnonInteraction2008,bomborHalfMetallicFerromagnetismUnexpectedly2013,madduriMagnoninducedInterbandSpinflip2017}, and possible unconventional superconductivity~\cite{erlandsenMagnonmediatedSuperconductivitySurface2020,thingstadEliashbergStudySuperconductivity2021,brekkeInterfacialMagnonmediatedSuperconductivity2024}. Yet, in stark contrast to EPC, EMC still lacks a quantitative first-principles description. The effect of EMC is significant and its absence in state-of-the-art first-principles calculations leads to clear failures. For example, first-principles calculations considering EPC-only largely underestimate the room-temperature resistivity of $\alpha$-Fe and yield a purely linear $T$ dependence without the experimentally observed $T^2$ component~\cite{alvarezElectronphononCouplingMagnetic2025}. This discrepancy has long been attributed to EMC, but without direct computational evidence~\cite{isshikiTemperatureDependenceElectrical1978,raquetMagneticResistivityElectron2002}.

Despite the broad significance of EMC, quantitative first-principles EMC calculations have been hindered by two distinct obstacles. First, the theoretical formalism remains incomplete. The many-body perturbation formalism for EMC and the formulas required to calculate the related transport properties have not been fully developed, not least because magnons carry spin angular momentum and therefore demand a treatment fundamentally different from lattice vibrations. Second, even when such expressions are available, the key quantity on which they depend, namely the response of the Kohn-Sham potential to spin deviations, remains inaccessible to conventional first-principles methods. Existing treatments either rely on simplified models with empirical parameters~\cite{raquetMagneticResistivityElectron2002} or are limited to zero temperature and relatively coarse Brillouin-zone sampling~\cite{mullerElectronmagnonScatteringElementary2019,nabokElectronPlasmonElectron2021,paischerNonlocalCorrelationEffects2023}. The latter, while applicable to quasiparticle spectra, cannot handle finite-temperature transport calculations. Thus, a first-principles theory for quantifying EMC effects in transport remains an open challenge.

Here we develop a unified formalism for EMC within many-body perturbation theory (MBPT), which is applicable to both collinear ferro- and antiferromagnets. By leveraging the machine-learning spinful Hamiltonian (MLSH) approach~\cite{zhongAcceleratingElectronicstructureCalculation2023}, our method overcomes the intractable computational challenge and enables quantitative EMC transport calculations at a fully \textit{ab initio} level for the first time. Our work bridges a fundamental gap in the quantitative description of transport in magnetic systems and provides a unified platform to treat EMC and EPC on an equal footing.

\section*{Theory and Methods}

We first derive the formalism within MBPT. In the low-excitation regime, the magnetic part of KS potential can be approximated by a first-order expansion:
\begin{equation}
    V^\text{KS, mag} \approx V^\text{KS, mag}_0
                     + \sum_{p\kappa} \nabla_{\mathbf{S}_{p\kappa}} V^\text{KS, mag} \cdot \Delta \mathbf{S}_{p\kappa},
\end{equation}
where $\mathbf{S}_{p\kappa}$ denotes the spin vector of atom $\kappa$ in unitcell $p$, and $\Delta \mathbf{S}_{p\kappa} = \mathbf{S}_{p\kappa} - \mathbf{S}^0_{p\kappa}$. The central quantity is therefore the gradient of the magnetic KS potential with respect to the spin vector. This term is difficult to obtain directly from conventional DFT, because local magnetic moments are determined by the spin-dependent electron density and cannot be treated straightforwardly as independent perturbation variables in DFPT. Moreover, finite-difference calculations are complicated by the difficulty of constraining both the directions and magnitudes of local magnetic moments to prescribed values. To overcome this difficulty, we adopt the machine-learning Hamiltonian method~\cite{zhongTransferableEquivariantGraph2023,zhongAcceleratingElectronicstructureCalculation2023}. We represent the magnetic KS potential as~\cite{zhongAcceleratingElectronicstructureCalculation2023}
\begin{equation}
    V^\text{KS, mag} = \sum_{p,\kappa} \Delta_{p\kappa} \mathbf{S}_{p\kappa} \cdot \boldsymbol{\sigma}_\text{Pauli},
\label{eq:VKSmag_apt}
\end{equation}
where $\boldsymbol{\sigma}_{\mathrm{Pauli}}$ denotes the vector of Pauli matrices. Since $\Delta_{p\kappa}$ is
independent of the spin orientation, the derivative in the linearized expansion is simply represented by it. And $\Delta_{p\kappa}$ can be directly extracted from the MLSH model trained by HamGNN through one-shot prediction~\cite{zhongAcceleratingElectronicstructureCalculation2023}.

\begin{figure}[htbp]
    \centering
    \includegraphics[width=1.0\linewidth]{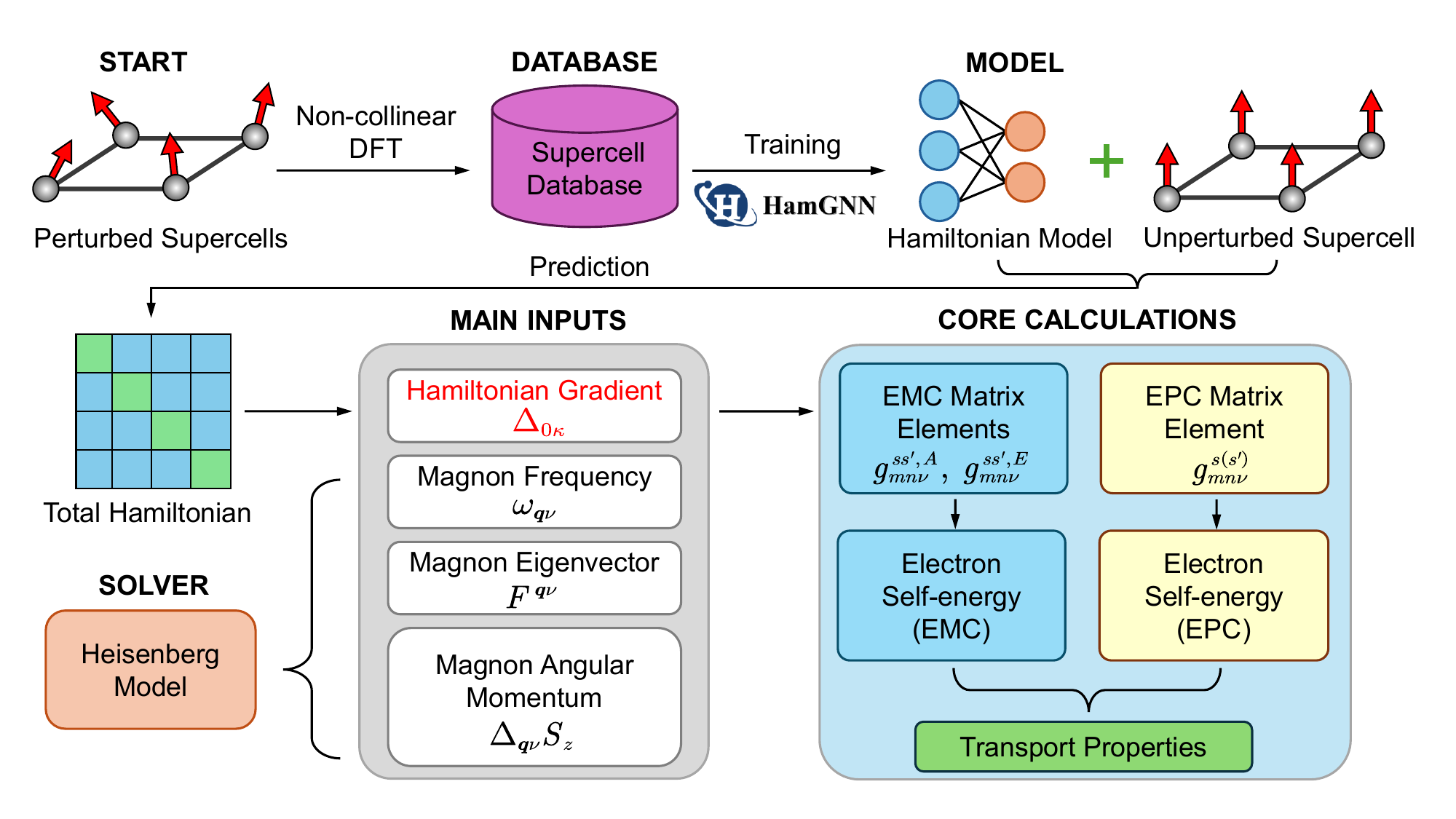}
    \caption{Framework for calculating EMC matrix elements and related transport properties. The initial step involves training the HamGNN Hamiltonian model based on non-collinear DFT calculation datasets to extract the Hamiltonian gradients $\Delta_{0\kappa}$. The eigenvalues $\omega_{\mathbf{q}\nu}$, eigenvectors $F^{\mathbf{q}\nu}$ and the magnon spin angular momentum $\Delta_{\mathbf{q}\nu} S_z$ are evaluated by solving the Heisenberg Hamiltonian. These quantities are combined with $\Delta_{0\kappa}$ to obtain the core EMC matrix elements, $g^{ss',\text{A}}_{mn\nu}$ and $g^{ss',\text{E}}_{mn\nu}$. Based on these primary inputs, the electron self-energy due to EMC is evaluated. Combining EMC self-energy result with that from EPC, the overall transport properties are calculated accordingly.}
\label{fig:frame}
\end{figure}

To quantize magnon excitations, we adopt the linear spin-wave theory (LSWT)~\cite{tothLinearSpinWave2015}, and focus on collinear magnetic structures with isotropic exchange interactions for now. Following the established diagonalization procedure~\cite{colpaDiagonalizationQuadraticBoson1978,petitNumericalSimulationsMagnetism2011,tothLinearSpinWave2015}, we solve the magnon eigenvalue problem and obtain the magnon frequencies $\omega_{\mathbf{q}\nu}$ together with the corresponding eigenvectors $\mathbf{F}^{\mathbf{q}\nu}$ [see Sec.~S1 in Supplemental Material (SM)~\cite{suppme} for detailed derivation].
\nocite{petitNumericalSimulationsMagnetism2011,tothLinearSpinWave2015,colpaDiagonalizationQuadraticBoson1978,okumaMagnonSpinMomentumLocking2017,alvarezElectronphononCouplingMagnetic2025,perdewGeneralizedGradientApproximation1996,kresseEfficiencyAbinitioTotal1996,kresseEfficientIterativeSchemes1996,kresseUltrasoftPseudopotentialsProjector1999,rosengaardFinitetemperatureStudyItinerant1997,togoFirstprinciplesPhononCalculations2023,togoImplementationStrategiesPhonopy2023,morrisonNonlocalHermitianNormconserving1993,ozakiNumericalAtomicBasis2004,ozakiVariationallyOptimizedAtomic2003,kurzInitioTreatmentNoncollinear2004,liechtensteinLocalSpinDensity1987,hanElectronicStructureMagnetic2004,terasawaEfficientAlgorithmBased2019,perdewAccurateSimpleAnalytic1992,perdewSelfinteractionCorrectionDensityfunctional1981,combescotCoherentEffectsPhysics2006,ryeeComparativeStudyDFT2018,ryeeEffectDoubleCounting2018,lamsalPersistenceLocalmomentAntiferromagnetic2013,ramazanogluRobustAntiferromagneticSpin2017,giustinoElectronphononInteractionsFirst2017,zhongAcceleratingCalculationElectron2024,mullerElectronmagnonScatteringElementary2019,laubitzTransportPropertiesFerromagnetic1976,whiteElectricalThermalResistivity1959,liuDirectMethodCalculating2015,
soulairolStructureMagnetismBulk2010,crangleMagnetizationPureIron1971,
pajdaInitioCalculationsExchange2001,
sunEffectiveDescriptorPredicting2025,
ponceFirstprinciplesCalculationsCharge2020,
claesAssessingQualityRelaxationtime2022,
loongNeutronScatteringStudy1984,mookTemperatureDependenceMagnetic1973}
To remove the ambiguity associated with degenerate magnon eigenvectors in AFM systems, we further derive a relation that pairs each adjoint mode with its corresponding normal mode. Introducing the magnon annihilation and creation operators ($\hat{\alpha}_{\mathbf{q}\nu}$ and $\hat{\alpha}^{\dagger}_{\mathbf{q}\nu}$), we obtain:
\begin{align}
    \Delta V^\text{KS, mag} &= \frac{1}{\sqrt{N}} \sum_{\mathbf{q}\nu}\sum_{p\kappa} \sqrt{\frac{S_\kappa}{2}} \Delta_{p\kappa} 
        \mathrm{e}^{i\mathbf{q}\cdot(\mathbf{R}_p+\boldsymbol{\tau}_\kappa)} \times \notag \\
            &\quad\quad\quad\quad\quad\quad\quad(D^\text{A}_{\kappa, \mathbf{q}\nu} \hat{\alpha}_{\mathbf{q}\nu} +
            D^\text{E}_{\kappa, \mathbf{q}\nu} \hat{\alpha}^\dagger_{-\mathbf{q}\nu}), \notag \\
    D^\text{A}_{\kappa, \mathbf{q}\nu} &= \begin{bmatrix}
        0 & 1-A_\kappa F^{\mathbf{q}\nu}_\kappa +  1+A_\kappa F^{\mathbf{q}\nu}_{\kappa+M}  \\
         1+A_\kappa F^{\mathbf{q}\nu}_\kappa + 1-A_\kappa F^{\mathbf{q}\nu}_{\kappa+M}  & 0
    \end{bmatrix}, \notag \\
    D^\text{E}_{\kappa, \mathbf{q}\nu} &= {(D^\text{A}_{\kappa, \mathbf{q}\nu})}^\mathrm{T},
\label{eq:deltaV_final}
\end{align}
where $A_\kappa = 1$ for spin-up atoms and $A_\kappa = -1$ for spin-down atoms. Using the standard second-quantization approach and taking momentum conservation into account, we obtain:
\begin{align}
    \hat{H}_\text{em} &= \frac{1}{\sqrt{N}} \sum_{nm}\sum_{\mathbf{k}\mathbf{q}}\sum_{ss'\nu}
        \hat{C}^\dagger_{m\mathbf{k+q}s'} \hat{C}_{n\mathbf{k}s}
        \left[
            g^{ss',\text{A}}_{mn\nu}(\mathbf{k},\mathbf{q}) \hat{\alpha}_{\mathbf{q}\nu} + 
            g^{ss',\text{E}}_{mn\nu}(\mathbf{k},\mathbf{q}) \hat{\alpha}^\dagger_{-\mathbf{q}\nu}
        \right] \notag \\
    g^{ss',\text{A/E}}_{mn\nu}(\mathbf{k},\mathbf{q}) &=
        \sum_\kappa \sqrt{\frac{S_\kappa}{2}} \mathrm{e}^{i\mathbf{q}\cdot\boldsymbol{\tau}_\kappa}
        (
            \sum_p \left\langle \psi_{m\mathbf{k+q}s'} \middle\vert \Delta_{p\kappa} \middle\vert \psi_{n\mathbf{k}s} \right\rangle
            \mathrm{e}^{i\mathbf{q}\cdot\mathbf{R}_p}
        )
        \left\langle s' \middle\vert D^\text{A/E}_{\kappa, \mathbf{q}\nu} \middle\vert s \right\rangle,
\end{align}
\noindent where $g^{ss',\text{A/E}}_{mn\nu}(\mathbf{k},\mathbf{q})$ is the EMC matrix element.
It should be emphasized that, since $D^\text{A/E}$ possesses only off-diagonal components, the first-order electron-magnon interaction inevitably induces an electron spin-flip process. Because magnons carry spin angular momentum, the angular-momentum conservation should be considered in electron-magnon interaction processes. The spin angular momentum of a magnon $\lvert\mathbf{q}\nu\rangle$ can be characterized by the deviation of $S_z$ upon excitations~\cite{okumaMagnonSpinMomentumLocking2017} (see Sec.~S2 in SM~\cite{suppme} for detailed derivations):
\begin{equation}
    \Delta_{\mathbf{q}\nu} S_z =
    \begin{cases}
        -1 & \text{spin-up-flip magnon},\,\uparrow \\
        +1 & \text{spin-down-flip magnon},\,\downarrow
    \end{cases}.
\end{equation}
According to angular-momentum conservation, the electron-magnon interaction can be classified into four distinct processes:
$\hat{C}^\dagger_\downarrow \hat{C}_\uparrow \hat{\alpha}_\uparrow,\,
\hat{C}^\dagger_\uparrow \hat{C}_\downarrow \hat{\alpha}^\dagger_\uparrow,\,
\hat{C}^\dagger_\uparrow \hat{C}_\downarrow \hat{\alpha}_\downarrow,\,
\hat{C}^\dagger_\downarrow \hat{C}_\uparrow \hat{\alpha}^\dagger_\downarrow$, and the first-order electron-magnon interaction Hamiltonian can be obtained correspondingly. With the interaction Hamiltonian, the formulas of electronic and transport properties can be derived. The overall framework is illustrated in Fig.~\ref{fig:frame} and the detailed derivations are shown in Sec.~S3 of SM~\cite{suppme}, which also contains the model-training and calculation details (Sec.~S4). We have also developed an atomic-orbital-based framework that efficiently and accurately calculates the EPC matrix elements for magnetic systems~\cite{zhongAcceleratingCalculationElectron2024} (see Sec.~S5 in SM~\cite{suppme}). The entire methodology has been implemented in our in-house code $\mathbf{QTrans}$~\cite{liElectronphononCouplingsDensity2026}.

\section*{Results of Ferromagnet}

\begin{figure}[htbp]
    \centering
    \includegraphics[width=0.8\linewidth]{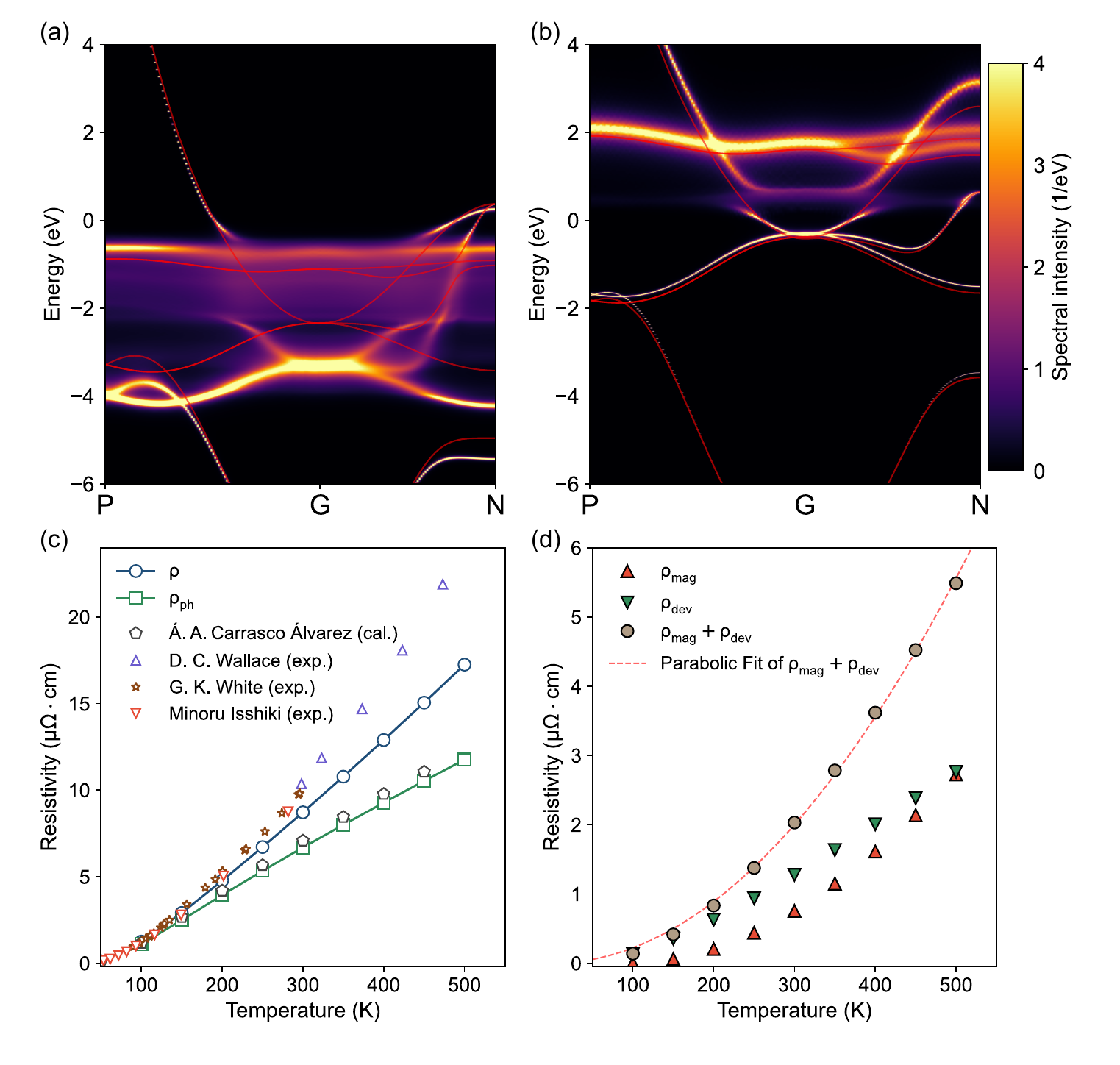}
    \caption{
        Calculated electron spectral functions and resistivity for $\alpha$-Fe.
        electron spectral functions induced by EMC for the (a) spin-up and (b) spin-down channels, respectively.
        (c) Calculated total resistivity ($\rho$) and the phonon-limited contribution ($\rho_\text{ph}$) as functions of temperature. 
        The experimental results are extracted from Refs.~\cite{wallaceSpecificHeatHigh1960,whiteElectricalThermalResistivity1959,isshikiTemperatureDependenceElectrical1978}, while the pentagon symbols denote the calculated $\rho_\text{ph}$ values reported in Ref.~\cite{alvarezElectronphononCouplingMagnetic2025}.
        (d) Calculated magnon-limited contribution ($\rho_\text{mag}$), the deviation from Matthiessen's rule ($\rho_\text{dev}$), and their combined contribution $\rho_\text{mag}+\rho_\text{dev}$. The dotted line represents a parabolic fit to $\rho_\text{mag}+\rho_\text{dev}$.}
\label{fig:Fe-results}
\end{figure}

To benchmark our framework, we begin to study the electronic structures and transport properties of
ferromagnetic metals $\alpha$-Fe and Ni. In the main text, we focus on $\alpha$-Fe as a representative ferromagnet, while the corresponding results for Ni are provided in Sec.~S6 of SM~\cite{suppme}. The details of the Heisenberg model are provided in Sec.~S7 in SM~\cite{suppme}. We first analyze how EPC and EMC modify quasiparticle properties through their respective contributions to the electron self-energy. The real part of self-energy leads to energy-level shifts, whereas the imaginary part results in quasiparticle lifetime broadening. Our calculated electron spectral functions for $\alpha$-Fe indicate that the renormalization due to EPC is relatively small (see Fig.~S6 in SM~\cite{suppme}). In contrast, EMC induces pronounced and spin-dependent band renormalization, underscoring its critical role in this FM system. For the spin-up channel, as shown in Fig.~\ref{fig:Fe-results}(a), the localized $d$-bands located approximately $1.0\,\text{eV}$ below the Fermi level shift upward by $\sim 0.3 \,\text{eV}$, while states around $-2.3\,\text{eV}$ shift downward by $\sim 1.0\,\text{eV}$. Within this energy window, the electronic states are strongly broadened, to the point of nearly losing their quasiparticle character. For the spin-down channel, however, the spectral peaks below the Fermi level remain sharp and well-defined, exhibiting only a modest upward shift relative to the DFT band dispersions [Fig.~\ref{fig:Fe-results}(b)]. A distinct feature emerges at $\sim 0.5\,\text{eV}$ above the Fermi energy, where the DFT band splits into two branches as a consequence of strong EMC. These results are consistent with the electron spectral functions reported in previous work~\cite{mullerElectronmagnonScatteringElementary2019} in the low-energy regime, where the interaction is primarily governed by collective magnon excitations. This agreement further validates our treatment of EMC.

The pronounced spectral broadening induced by EMC clearly demonstrates its strong influence on quasiparticle lifetimes. Therefore, we proceed to systematically investigate the electrical resistivity, aiming to quantitatively assess the individual and combined contributions of both EPC and EMC. The calculated temperature-dependent resistivity of $\alpha$-Fe is shown in Fig.~\ref{fig:Fe-results}(c,d). To distinguish and quantify the individual impacts of EMC and EPC, we decompose the total resistivity $\rho$ into three distinct components:
\begin{equation}
    \rho = \rho_\text{ph} + \rho_\text{mag} + \rho_\text{dev},
\label{eq:rho_terms}
\end{equation}
where $\rho_\text{ph}$ and $\rho_\text{mag}$ denote the resistivity arising solely from EPC and EMC, respectively. The deviation term $\rho_\text{dev}$, quantifies the deviation from the Matthiessen's rule~\cite{zimanElectronsPhononsTheory1960,bassDeviationsMatthiessensRule1972,raquetElectronmagnonScatteringMagnetic2002,glasbrennerDeviationsMatthiessensRule2014,leveilleeInitioCalculationCarrier2023}, originating from the interplay between electron-phonon and electron-magnon scattering processes. The calculated $\rho_\text{ph}$ is consistent with a recent theoretical study based on DFPT~\cite{alvarezElectronphononCouplingMagnetic2025}, validating the accuracy of our atomic-orbital-based EPC approach. Above $100\,\text{K}$, $\rho_\text{ph}$ exhibits an approximately linear temperature dependence, consistent with the Gr\"uneisen model~\cite{isshikiTemperatureDependenceElectrical1978}. However, such a linear behavior fails to reproduce the $T^2$ contribution observed experimentally~\cite{isshikiTemperatureDependenceElectrical1978,volkenshteinScatteringMechanismsConduction1973}. Moreover, its magnitude ($6.7\,\mu\Omega\cdot\text{cm}$ at $300\,\text{K}$) accounts for about $70\%$ of the reported experimental values of $\sim 10 \,\mu\Omega\cdot\text{cm}$. After taking EMC into account, the discrepancy is largely reconciled. The total resistivity, comprising all three contributions in Eq.~\ref{eq:rho_terms}, reaches $8.7\,\mu\Omega\cdot\text{cm}$ at $300\,\text{K}$, yielding substantially improved agreement with experiment. The possible reasons for the residual discrepancies are discussed in Appendix~A.

As shown in Fig.~\ref{fig:Fe-results}(d), the enhanced resistivity originates from the dual role of EMC. The first contribution is the direct magnon-scattering term, $\rho_\text{mag}$, which has been anticipated in previous studies~\cite{volkenshteinScatteringMechanismsConduction1973,alvarezElectronphononCouplingMagnetic2025}. The second contribution, denoted as $\rho_\text{dev}$, quantifies the deviation from Matthiessen's rule. Contrary to the common assumption that this term is negligible~\cite{raquetElectronmagnonScatteringMagnetic2002}, we find that $\rho_\text{dev}$ exceeds $\rho_\text{mag}$ in $\alpha$-Fe. Physically, this term originates from the fact that magnon scattering is a spin-flip process that couples the two spin channels, whereas phonon scattering is spin-conserving within each channel. In the exchange-split bands of $\alpha$-Fe, the two mechanisms therefore cannot be treated as independent additive scattering channels, and the strong spin asymmetry of $\alpha$-Fe amplifies the resulting deviation term (see Sec.~S8 in SM~\cite{suppme} for additional discussion). Thus, the $T^2$ component of the resistivity is not generated by EMC alone, as has long been assumed, but is instead governed by the interplay between EPC and EMC. Remarkably, the composite contribution $\rho_\text{mag} + \rho_\text{dev}$ exhibits an excellent $T^2$ temperature dependence, with a coefficient of $2.2 \times 10^{-5}\,\mu\Omega\cdot\text{cm}\cdot\text{K}^{-2}$, in good agreement with the experimental result of $2.2 \pm 0.1 \times 10^{-5}\,\mu\Omega\cdot\text{cm}\cdot\text{K}^{-2}$~\cite{isshikiTemperatureDependenceElectrical1978}. Consequently, the remaining subtle discrepancy relative to the measurements should exhibit a linear temperature dependence. These results establish that EMC, particularly through its interplay with EPC, is indispensable to charge transport in $\alpha$-Fe, and refine the long-standing attribution of the $T^2$ resistivity solely to single-magnon scattering. Note that EPC-EMC interplay is even more dominant in Ni (see Sec.~S6 in SM~\cite{suppme}).

We also evaluate the dimensionless EMC strength $\lambda_\text{EMC}$ of $\alpha$-Fe and obtain a value of 1.66, substantially larger than the EPC strength, $\lambda_\text{EPC} = 0.34$. Nevertheless, EMC does not dominate the total resistivity of $\alpha$-Fe. This discrepancy arises from the very different energy scales of the two bosonic quasiparticles. Magnons in $\alpha$-Fe extend up to $\sim 500\,\text{meV}$, more than an order of magnitude above the phonon cutoff energy of $\sim 40\,\text{meV}$ (see Fig.~S3 in SM~\cite{suppme}), and are therefore much less thermally populated at room temperature. As a result, the direct magnon-limited resistivity $\rho_\text{mag}$ remains smaller than the phonon contribution. This naturally raises the question of whether EMC can dominate charge transport outright. To address this question and demonstrate the generality of our framework, we turn to the antiferromagnet $\text{Ba}_{1-x}\text{K}_x\text{Mn}_2\text{As}_2$.

\begin{figure}[htbp]
    \centering
    \includegraphics[width=1.0\linewidth]{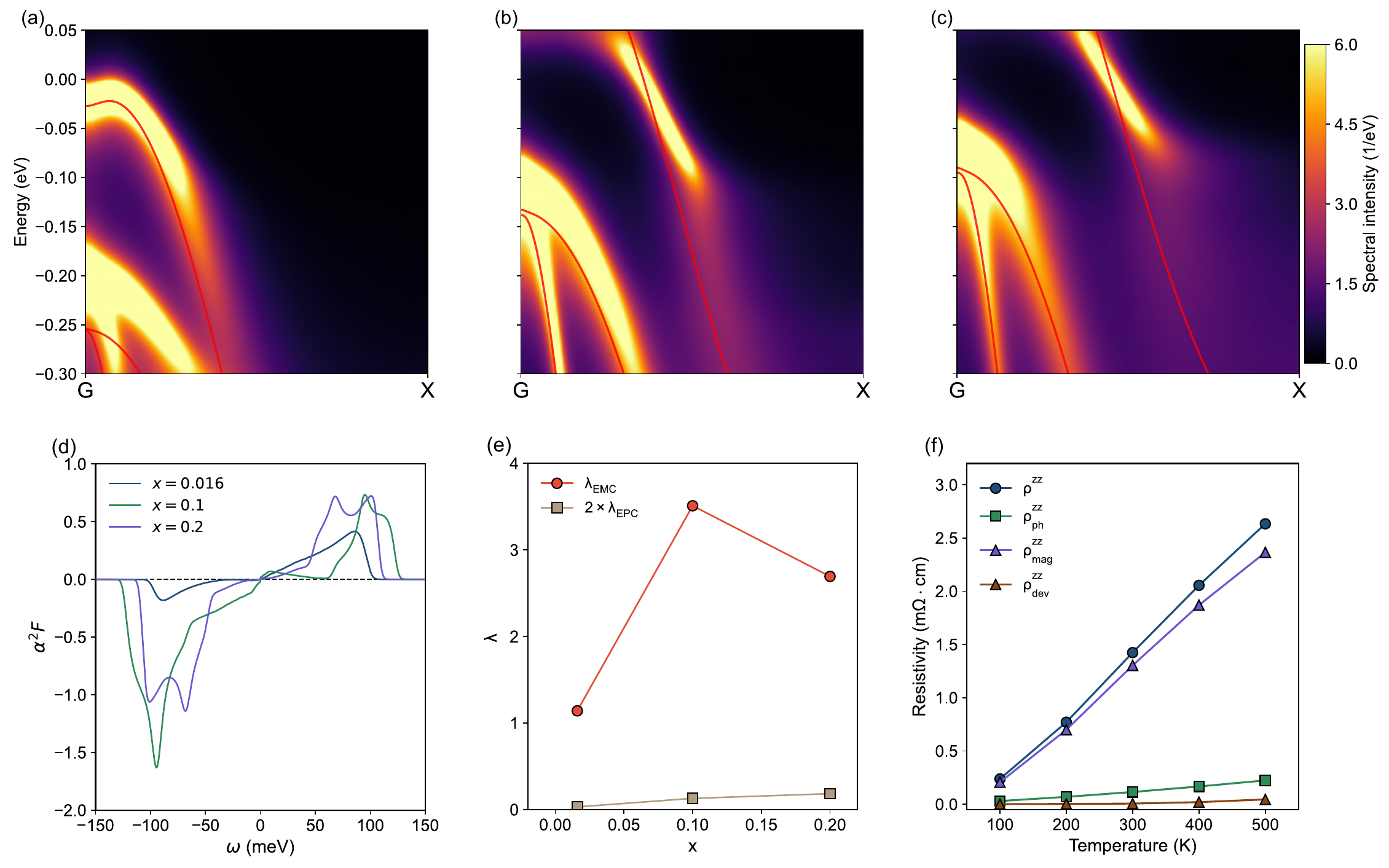}
    \caption{
        Calculation results for $\text{Ba}_{1-x}\text{K}_x\text{Mn}_2\text{As}_2$.
        Electron spectral functions induced by EMC for (a) $x=0.0$, (b) $x=0.1$ and (c) $x=0.2$, respectively.
        The red lines represent the DFT band structures.
        (d) Calculated Eliashberg spectra for different doping levels. Unlike the phonon case, magnon absorption and emission processes involve distinct EMC matrix elements. Accordingly, the positive (negative) part of the spectrum is associated with magnon absorption (emission).
        (e) Calculated EMC strength $\lambda_\text{EMC}$ (circles) and EPC strength $\lambda_\text{EPC}$ (squares) as functions of doping level.
        According to the definition in previous work~\cite{alvarezElectronphononCouplingMagnetic2025}, the EPC strengths are multiplied by 2 for comparison.
        (f) Calculated total resistivity ($\rho$), the phonon-limited contribution ($\rho_\text{ph}$), magnon-limited contribution ($\rho_\text{mag}$) and the deviation from Matthiessen's rule ($\rho_\text{dev}$) as functions of temperature for charge transport along $z$ direction.}
\label{fig:BMA-results}
\end{figure}

\section*{Results of Antiferromagnet}

$\text{Ba}_{1-x}\text{K}_x\text{Mn}_2\text{As}_2$ is a local-moment antiferromagnetic metal closely related to the 122-type iron-based superconductors~\cite{christiansonUnconventionalSuperconductivityBa06K04Fe2As22008,yuStrongBandRenormalization2022}. The calculated magnetic moments agree with experiments across the doping series (see Table~S3 in SM~\cite{suppme}). Using exchange parameters extracted from the experiment~\cite{ramazanogluRobustAntiferromagneticSpin2017}, we obtain the corresponding magnon dispersions (see Fig.~S5 in SM~\cite{suppme}). The EMC-induced spectral functions [Fig.~\ref{fig:BMA-results}(a-c)] display clear kinks at $\sim 100\,\text{meV}$ below the Fermi level that sharpen with doping, in good agreement with ARPES measurements~\cite{yuStrongBandRenormalization2022}. This agreement validates our framework in this very different class of magnet.

We next quantify the EMC strength in doped $\mathrm{BaMn_2As_2}$. The Eliashberg spectral function $\alpha^2F(\omega)$ and the dimensionless EMC coupling strength $\lambda_\text{EMC}$ [Fig.~\ref{fig:BMA-results}(d,e)] are remarkably large. Specifically, $\lambda_\text{EMC}$ rises from 1.14 at $x=0.016$ to a peak value of 3.51 at $x=0.1$, before decreasing to 2.69 at higher doping level. These values are of the same order as the ARPES estimate~\cite{yuStrongBandRenormalization2022} and are more than an order of magnitude larger than $\lambda_\text{EPC}$. For instance, at $x=0.016$, $\lambda_\text{EMC}=1.14$, whereas $\lambda_\text{EPC}=0.03$. These results establish EMC, rather than EPC, as the dominant electron-boson coupling shaping the low-energy electronic properties.

To uncover the microscopic origin of the EMC dominance, we decompose $\lambda$ using the frequency-dependent nesting function proposed by \textit{Sun et al.}~\cite{sunEffectiveDescriptorPredicting2025} (see Sec.~S10 in SM~\cite{suppme}). This decomposition reveals a counterintuitive mechanism. Magnons provide more than an order of magnitude fewer scattering channels than phonons, but their matrix elements are roughly an order of magnitude larger, owing to the strong exchange interactions in this system~\cite{anElectronicStructureMagnetism2009}. Since $\lambda$ scales with the squared matrix element, the dominance of EMC over EPC is thus governed by the matrix elements rather than the available phase space. A mode-resolved analysis (see Fig.~S10 in SM~\cite{suppme}) further shows that the dominant contribution shifts from $\Gamma$-point magnons at low doping to finite-$\mathbf{q}$ magnons at higher doping.

We then assess whether this EMC dominance also manifests in charge transport. As shown in Fig.~\ref{fig:BMA-results}(f), we calculate the resistivity of $\text{Ba}_{1-x}\text{K}_x\text{Mn}_2\text{As}_2$ at a doping level of $x = 0.016$ within the rigid-band approximation (see Sec.~S11 in SM~\cite{suppme} for the resistivity along the $x$ direction). Notably, the magnon-limited resistivity $\rho_\text{mag}$ is approximately 6--11 times as large as the phonon-limited contribution $\rho_\text{ph}$. This behavior contrasts with $\alpha$-Fe, where the high characteristic energy scale of the magnon spectrum keeps the direct magnon contribution below the phonon contribution. In $\text{Ba}_{1-x}\text{K}_x\text{Mn}_2\text{As}_2$, the EPC is negligible ($\lambda_\text{EPC}=0.03$), allowing EMC to dominate the resistivity outright. The overall resistivity along the $z$ direction $\rho_\text{zz}$ as well as the magnon-limited contribution increases linearly with temperature, similar to the phonon scattering behaviour~\cite{amarelExactSolutionBoltzmann2020}. We further find that the deviation term $\rho_\text{dev}$ is unexpectedly small in this AFM system, indicating that Matthiessen's rule is approximately satisfied. These results reveal that the nature of magnetic ordering dictates not only the magnitude of spin-dependent scattering, but also the degree to which distinct scattering mechanisms intertwine (see Sec. S12 for detailed discussion).

\section*{Discussion and Conclusion}

Across the two systems, a unified picture emerges: $\lambda_\text{EMC}$ substantially exceeds $\lambda_\text{EPC}$ in both the itinerant ferromagnet $\alpha$-Fe (1.66 versus 0.34) and the local-moment antiferromagnet $\text{BaMn}_2\text{As}_2$ (1.14 versus 0.03). This suggests that EMC can be the dominant electron-boson coupling in magnetic metals, even though its consequences for transport differ between the two.

In summary, we have established a unified first-principles framework for EMC in collinear magnets, combining a MLSH with LSWT within MBPT. Our framework provides quantitative access to a full set of EMC observables, including quasiparticle spectra, the dimensionless coupling strength, and transport. We validate it in the ferromagnet $\alpha$-Fe and the antiferromagnetic $\text{Ba}_{1-x}\text{K}_x\text{Mn}_2\text{As}_2$. Our framework resolves the long-standing $T^2$ resistivity contribution of ferromagnets, which we find to arise not from single-magnon scattering alone but also from the interplay of electron-phonon and electron-magnon scattering. Our results also suggest EMC as the dominant electron-boson coupling in these magnetic metals in terms of coupling strength. By making EMC computable from first principles, our approach enables quantitative studies of transport, spectroscopy, and many-body effects across magnetic materials, with natural extensions to higher-order scattering processes, and to magnon-mediated phenomena such as possible unconventional superconductivity.

\begin{acknowledgments}
This work was supported by the National Natural Science Foundation of China (12188101), the China National Key Research and Development Program (2022YFA1404603), the National Natural Science Foundation of China (12474222), Shanghai Science and Technology Program (23JC1400903), Quantum Science and Technology-National Science and Technology Major Project (2024ZD0300102), and the Guangdong Major Project of the Basic and Applied Basic Research (2021B0301030005).
\end{acknowledgments}

\vspace{0.5cm}

\section*{Author contribution}

J.-H.Y. conceived and supervised the project. H.X. and X.-G.G. co-supervised the project. S.L. developed the theoretical formalism, and H.L. derived the IBTE formalism. X.L. and H.L. checked the formalism. S.L., X.L. and H.L. implemented the codes, performed the calculations and prepared the manuscript. Y.Z. provided the interface of HamGNN. J.-H.Y., H.X., and X.-G.G. revised the manuscript.

\bibliography{main}

\appendix

\section{Residual Discrepancy in Resistivity}

The calculated resistivities of $\alpha$-Fe and Ni are in good agreement with experimental data, yet a noticeable discrepancy remains. This residual discrepancy may originate from physical effects not fully captured within the present framework, including, for example, the renormalization of the magnetic moment~\cite{liuDirectMethodCalculating2015} or the neglect of higher-order Feynman diagrams~\cite{zhouPredictingChargeTransport2019,luoFirstprinciplesDiagrammaticMonte2025,lihmBeyondQuasiparticleTransportVertex2026}. Additionally, the present analysis of the interplay between electron-phonon and electron-magnon scattering relies on the self-energy relaxation time approximation. More advanced treatments, such as the iterative Boltzmann transport equation method, may introduce additional coupling to these two scattering sources (see Sec.~S12 in SM~\cite{suppme}). Furthermore, higher-order diagrammatic contributions, as well as explicit magnon-phonon interactions, could provide additional scattering pathways beyond the scope of this work. Nevertheless, the present framework enables an efficient and first-principles evaluation of EMC matrix elements. This capability not only allows for the simultaneous inclusion of EPC and EMC effects in practical calculations, but also establishes a solid foundation for extensions within the MBPT framework, even for non-perturbative treatments~\cite{luoFirstprinciplesDiagrammaticMonte2025,lihmBeyondQuasiparticleTransportVertex2026}.

\end{document}